\documentclass[aps,pra,onecolumn,amssymb]{revtex4}
\input epsf.tex

\begin{document}

\title{Quantum Relays for Long Distance Quantum Cryptography}

\author{Daniel Collins, Nicolas Gisin and Hugues de Riedmatten}
\affiliation{Group of Applied Physics, University of Geneva, 20,
rue de l'Ecole-de-M\'{e}decine, CH-1211 Geneva 4, Switzerland}

\date{14 November 2003}

\begin{abstract}

Quantum Cryptography is on the verge of commercial application.
One of its greatest limitations is over long distance - secret key
rates are low and the longest fibre over which any key has been
exchanged is currently 100 km. We investigate the quantum relay,
which can increase the maximum distance at which quantum
cryptography is possible.  The relay splits the channel into
sections, and sends a different photon across each section,
increasing the signal to noise ratio.  The photons are linked as
in teleportation, with entangled photon pairs and Bell
measurements. We show that such a scheme could allow cryptography
over hundreds of kilometers with today's detectors.  It could not,
however, improve the rate of key exchange over distances where the
standard single section scheme already works.  We also show that
reverse key reconciliation, previously used in continuous variable
quantum cryptography, gives a secure key over longer distances
than forward key reconciliation.

\end{abstract}

\maketitle

\newcommand{\ket}[1]{\left | #1 \right \rangle}
\newcommand{\bra}[1]{\left \langle #1 \right |}
\newcommand{\amp}[2]{\left \langle #1 \left | #2 \right. \right \rangle}

\vspace{0.2cm}

\section{Introduction}

Digital security of transmitted information forms a key part of
internet shopping, internet banking, and remote access to
corporate networks.  Since no cryptosystem is perfect, one would
like to use the most secure system possible.  Quantum cryptography
\cite{Gisin1} promises to be a key part of that system.  Its
strength is that it is based upon a fundamentally different
assumption to all currently available systems.  While they are
based upon our knowledge of the time required to perform certain
computations, quantum cryptography is based upon our knowledge of
the laws of physics, and in particular quantum mechanics.

Quantum cryptography is currently in the development stage. One
implementation is to send bits as time-bin single photons down a
fibre optic cable, and to use quantum complementarity to detect
the presence of any eavesdropper.  Many experiments have shown its
practicality, including recently over a physical distance of 67km
\cite{67km}, and over 100km of fibre in the laboratory
\cite{100km}.

A limitation to the development of quantum cryptography is its
performance over distance.  Two effects come into play here. The
first is the exponential loss of signal, and hence bit rate, with
increasing distance. The second is the fundamental insecurity of
the quantum cryptosystem at long distances due to a roughly
constant level of noise dominating the exponentially decreasing
signal.  Both these obstacles can be overcome in theory using a
quantum repeater \cite{repeater}.  This is the idea of breaking up
the channel into several short sections and creating a perfect
entangled pair spanning each section, before using entanglement
swapping to make a single entangled pair spanning the whole
channel. The entangled pair can then be used to make a secret key,
using for example the Ekert protocol \cite{Ekert}. In order to
make a perfect entangled pair across a single section, we send
many pairs, each of which arrives with a small amount of noise,
and possibly not at all.  We then use a quantum memory,
non-demolition measurement, and entanglement purification
\cite{purification} to make one perfect entangled pair.

The main problem with such a scheme is precisely that it requires
a complicated set of quantum operations, a quantum memory and a
photon non-demolition measurement.  All these seem impractical in
the near future.

A simpler scheme which is on the verge of being implemented is the
quantum relay \cite{Gisin1, Gisin2, Yamamoto, Franson, 222}.  This
works in the same way as the quantum repeater, only without the
entanglement purification, the quantum memory, or the
non-demolition measurement. It no longer combats the exponential
loss of signal with distance, but does increase the maximum
distance over which secure quantum key distribution is possible.
This is achieved through an increase in the signal to noise ratio
at each detector, since each photon only has to pass some fraction
of the total channel.

For us a quantum "relay" with a single section is the standard
BB84 \cite{BB84} protocol where we send a photon from Alice to
Bob.  Two sections means we put a source of entangled photon pairs
in the middle of the channel, and send one photon of each pair to
Alice, and one to Bob.  We then use the BBM92 \cite{BBM92}
protocol to make the key.  Three sections means we perform
teleportation \cite{teleportation}, with a single photon source
with Alice, an entangled pair source two-thirds of the way from
Alice to Bob, and a Bell measurement one-third of the way from
Alice to Bob. With teleportation one then uses the BB84 protocol.
Four sections means we use entanglement swapping, with two
entangled sources $\frac{1}{4}$ and $\frac{3}{4}$ of the way down
the channel, and a Bell measurement half way down the channel. One
can use the BBM92 protocol.  For more sections, we use the natural
extension of these ideas.

We here study the effectiveness of the quantum relay for quantum
key distribution, under certain assumptions on the type of
eavesdropping which may be performed.  We shall show that the
relay could be useful for small numbers of sections.  However when
many sections are introduced, the extra noise they introduce
overwhelms their advantages, limiting the maximum achievable
distance.

One of the strongest proofs of quantum cryptography is by Inamori,
L\"{u}tkenhaus and Mayers \cite{Inamori}, and considers an
imperfect source, detectors, and channel, and allows Eve to
perform any attack on the channel, including a coherent attack
upon many successive qubits.  An elaboration is to even allow Eve
some limited undetected control over Alice and Bob's laboratories,
and to show that even then the protocol is still secure
\cite{invading}.  Here we consider a simplified scenario which
contains the main idea of the quantum repeater and is nevertheless
not too far from reasonable physical assumptions upon the limit of
Eve's technology.

We assume that all the sources emit a single photon (or a perfect
entangled pair of photons) every time, as opposed to being weak
coherent pulses, or entangled pair sources which sometimes emit
two pairs of photons, and often no pairs at all. Thus Eve has no
possibility for the photon number splitting (PNS) attack
\cite{PNS}.  With a more realistic source which sometimes sends
multi-photon pulses Eve could use the PNS attack to eavesdrop upon
Alice's source. However there is a protocol which reduces the
effectiveness of such eavesdropping \cite{SARG}. And for even
numbers of sections, we only have entangled photon pair sources,
and the PNS attack does not help Eve at all \cite{Gisin1}.

Whilst including the PNS attack would not dramatically alter our
results, a bigger effect would come from the loss of signal due to
such imperfect sources.  Typically a weak coherent pulse (or an
entangled pair source) only sends a photon (pair) a fraction $0.1$
of the time, and thus would reduce the signal by a factor $0.1$.
If we have $m$ sources, the signal between Alice and Bob will be
reduced by a factor $0.1^m$.  This in turn would reduce the
maximum distance at which we can create a key, by around $40 m$
km.  In addition it will reduce the key rate per pulse sent by a
factor $0.1$.  This becomes important since we can at most pulse
our source at around $10$ GHz, and at long distances most of the
photons will be lost, leading to low secret key rates.

Entangled sources which sometimes emit two pairs of photons also
introduce noise, since Alice may detect only the photon from one
pair, and Bob only the photon from the other pair.  Treating such
noise would complicate our calculations, without changing the main
message of the paper - that relays increase the possible distance
of quantum cryptography - and so we ignore it by assuming perfect
sources.

We shall assume that the channel has state-independent losses,
that the detectors are inefficient, that they have dark counts,
and that there is some imperfect optical visibility in the
channel. This shall allow us to calculate the signal (and noise)
between Alice and Bob.

The idea of quantum cryptography is that Eve can learn nothing
without introducing errors into the channel. For eavesdropping
purposes we assume that Eve has improved the optical components
which lie outside the laboratories of Alice and Bob, and that all
the noise we had thought due to those components is actually due
to Eve's cloning attack. Thus the noise due to optical
imperfections and dark counts outside Alice's and Bob's labs can
be due to the eavesdropper, but the dark counts inside Alice and
Bob's labs cannot.  This last point is in contrast to previous
studies of the quantum relay, in which all the noise, regardless
of its source, is considered due to the eavesdropper.

Since the last point has caused some debate, we shall review the
assumptions which go into cryptography.  One assumes that Alice
and Bob are in two separate offices, and that within each office
Eve has no capability whatsoever. Eve may perform any operation
allowed by the laws of physics anywhere outside the offices. There
are two communication channels connecting the two offices, one
classical and one quantum.  The first assumption, that Eve cannot
modify or hear anything which occurs inside Alice and Bob's
offices, is impossible to verify in practice.  However closely
they look, Eve could always have planted a device inside their
offices which is telling her everything they do. All that they can
do is to search their offices as best they can for such a device,
and so long as none is found to assume that none exists. Similarly
with the classical and quantum channels, we check that the sending
and receiving equipment located in our offices is functioning as
we expect, and then assume that they continue to behave in this
way. In particular, any dark counts inside our office are a
consequence of the normal working of the receiver, and so are out
of the reach of Eve. Whilst we must check that our detector
functions correctly, this is not a new assumption: it was always
there in classical cryptography!  Of course, any detectors outside
the labs of Alice and Bob are under Eve's control, and dark counts
belonging to such detectors can be used for eavesdropping.

A related issue is the detector efficiency.  If we increase the
detector efficiency, the signal to noise ratio will increase, just
as it did when we reduced the dark counts.  Instead of altering
the dark counts, Eve could alter the detector efficiency to gain
some information.  As with the dark counts, she can alter the
efficiency of detectors outside Alice and Bob's labs, but not
inside.  She could influence the count rates inside Alice and
Bob's labs by increasing the efficiency of the source, but they
will detect this if they look at coincidence count rates. Looking
at the information Eve learns about the bits sent, her possible
alterations of the detector efficiency are equivalent to her
possible alterations of the dark count rate, and so we analyze
only her dark count eavesdropping from now on.

We limit Eve to performing the optimal individual attack on single
photons, ie. cloning \cite{individualattack} the qubit, and then
measuring the clone after the bases are publicly announced. Whilst
in principle she could perform quantum coherent attacks involving
many successive qubits at once, we know of no simple way for her
to use such an attack to her advantage. In order to reduce the
mathematical complexity and focus upon our main task we ignore
them.

To put our work into context, one of us (N.G.) co-authored two
papers which included calculations for small numbers of relays for
a scenario similar to the one here, but which allows Eve to use
all the noise, including dark counts in Alice and Bob's labs, for
her eavesdropping \cite{Gisin1, Gisin2}. Waks, Zeevi and Yamamoto
considered splitting the channel into several sections, each of
which has an imperfect EPR source in the middle, and using
entanglement swapping to create the entanglement between Alice and
Bob and hence the key \cite{Yamamoto}.  Their scheme has the
advantage that the PNS attack does not help Eve for such systems.
They did not consider the teleportation-like schemes which we
shall consider here. Jacobs, Pittman and Franson considered a
scheme in which a signal is sent down the channel, and at various
intervals it is "relayed" using an EPR source and a Bell
measurement located next to one another \cite{Franson}.  Indeed,
they introduced the relay terminology.  They consider attenuation
in the channel and dark counts in the detectors. Their scheme is
closest in conception to a classical relay, but can be improved by
spreading out the EPR sources and detectors down the channel, as
we shall do here.

\section{Why distance is a problem}

To analyse the key rate of the quantum relay under our specified
conditions, we shall start with a single section of channel: ie.
having no relay at all.  We consider the standard BB84 protocol
\cite{BB84} in which Alice first chooses a basis at random, either
$ \{ \ket{0}, \ket{1} \} $, or $ \{
\frac{1}{\sqrt{2}}(\ket{0}+\ket{1}),
\frac{1}{\sqrt{2}}(\ket{0}-\ket{1}) \}$.  She then sends a random
bit $0/1$ encoded as the first/second basis vector. Bob measures
in one of the two basis, choosing at random.  Alice and Bob then
publicly announce the bases in which they sent/measured the
photon, using an authenticated public classical channel. They
accept the result if they used the same basis.

A further rule is that if Bob gets clicks in both of his detectors
simultaneously, the result is rejected.  This is standard in our
situation with perfect sources, but throws away useful information
for countering the PNS attack when we are using weak laser pulses.

The bits which remain are collectively called the sifted key.
Alice and Bob then perform error correction and privacy
amplification to get a perfect shared secret key, about which Eve
has zero knowledge.

To model the channel, we suppose there are losses of $\alpha$
dB/km, ie. that the probability the photon passes through $d$ km
of fibre is $t= 10^{-\alpha d /10}$.   The optical visibility,
which is the visibility one would obtain between Alice and Bob
with perfect source and detectors, is denoted $V_{opt}$.  One can
view this as that with probability $V_{opt}$, the optical part of
the channel works perfectly, and with probability $1-V_{opt}$, the
channel gives white noise (an equal mixture of $0$ and $1$). Since
most of the optical imperfections are due to imperfect detectors,
we shall assume that this is constant, and does not increase with
distance. The probability of having a dark count in any one
detector is D, and the probability of detecting the photon when it
arrives is $\eta$. As stated before, we assume a perfect single
photon source.

In order for Bob to correctly receive and accept the (noiseless)
signal sent by Alice, the photon must pass the fibre unabsorbed
($t$) and without becoming noise ($V_{opt}$), and be detected in
Bob's detector ($\eta$).  Bob must measure in the correct basis
($\frac{1}{2}$), and without a dark count in the wrong detector
($1-D$).  This gives the probability
\begin{equation}
P(signal) = \frac{1}{2} t \eta V_{opt} (1-D).
\end{equation}

For Bob to accept a bit for the sifted key, he must measure in the
correct basis ($\frac{1}{2}$).  Either a photon must pass the
fibre ($t$) and be detected ($\eta$), or no photon must be
detected ($1-t\eta$) but instead a dark count in either detector
($2D$).  There must also be no dark count in the other detector
($1-D$). This gives a total probability
\begin{equation}
P(total) = \frac{1}{2}(t \eta + (1-t \eta) 2 D) (1-D).
\end{equation}

The noise is given by $P(noise)=P(total)-P(signal)$.

Bob's visibility of Alice's Bit, $V_{AB}$, is given by
\begin{eqnarray}
V_{AB} & = & \frac{P(signal)}{P(total)} \\
& = &  \frac{t \eta V_{opt}}{t \eta + (1-t \eta) 2 D}.
\end{eqnarray}

The probability that, when Bob detects something, Alice and Bob's
bits of the sifted key agree is
\begin{equation}
P(A=B)  =  \frac{1}{2} (1+V_{AB}).
\end{equation}

Eve's knowledge about Alice's bit comes from replacing the
imperfect optical visibility of the channel with perfect optical
visibility, and performing the optimal individual qubit attack
which introduces an amount of noise equal to $V_{opt}$. This was
proven in \cite{individualattack} to be a cloning attack. The
relationship between Eve's visibility of Alice's bit ($V_{AE}$)
and the visibility she may use for eavesdropping ($V_{opt}$) is
given by putting together equations (64, 7 and 67) of that paper,
resulting in the simple relation
\begin{equation}
V_{AE} = \sqrt{1-V_{opt}^2}.
\end{equation}

\begin{figure}[h]
\begin{center}
\epsfxsize=9cm \epsfbox{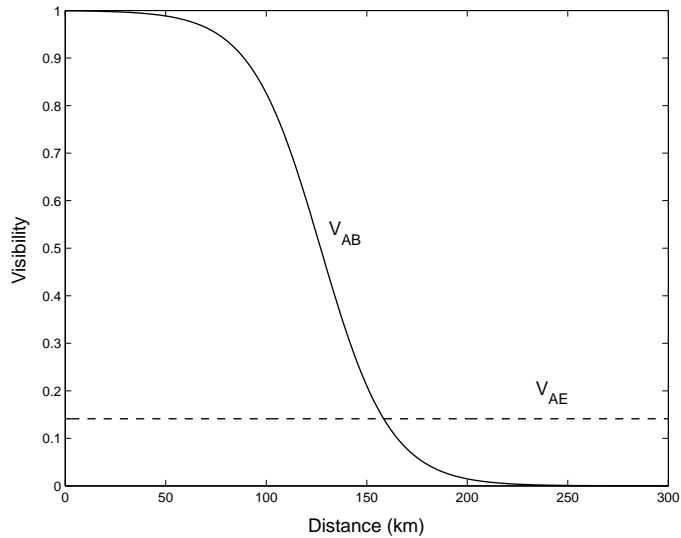} \caption{The visibility of
Alice's bit with increasing distance.} \label{onelegV}
\end{center}
\end{figure}

Fig \ref{onelegV} shows the visibility of Alice's bit for Bob
(solid line) and for Eve (dashed line) with increasing distance.
Here and elsewhere in this paper (unless otherwise stated), we
take realistic experimental parameters for a photon of wavelength
$1550 nm$ of $\alpha=0.25 dB/km$, $\eta=0.3$, $D=10^{-4}$, and
$V_{opt}=0.99$.  Section \ref{darkvefficiency} contains a
discussion of the possible tradeoff between dark counts and
detector efficiency.

At zero distance, Bob has a visibility of almost $0.99$. This is
because dark counts are negligible, and so the visibility is
approximately $V_{opt}$.  As the distance increases, more and more
of the photons are absorbed whilst the dark counts remain
constant, giving a lower signal to noise ratio and so worse
visibility. Eve's visibility remains constant, since we have
assumed that she can only use the optical visibility for
eavesdropping, and this we assume constant.  When Eve's visibility
is better than Bob's, Alice and Bob can no longer make a secret
key using one-way communication from Alice to Bob.

It looks like we could go to $150$ km or so, but this is partly
due to our perfect single photon source.  With a more realistic
source which only emits a photon with probability $0.1$, the
signal will be diminished by $10$ dB.  A $10$ dB loss is like $40$
km of fibre, and so we lose $40$ km, giving us a maximum distance
around 110 km (ignoring photon number splitting attacks).

Notice that one can tolerate a very noisy signal, and still distil
a secret key.  This seems in contrast to various claims that the
visibility must be above $\frac{1}{\sqrt{2}}=0.707$, or
equivalently that the fidelity ($P(A=B)$) must be above $0.854$.
The difference is that here we do not allow Eve to use the dark
counts of the eavesdropper, whereas some other calculations do.
This difference highlights the importance of good assumptions.

After error correction and privacy amplification, which we will
assume performed optimally in the asymptotic limit with one way
communication from Alice to Bob \cite{maurer}, they share a
perfectly secure key with a rate (per qubit sent) of
\begin{eqnarray}
Rate_{AB} & = & P(total) (I_{AB} - I_{AE}) \\
& = & \frac{1}{2}(t \eta + (1-t \eta) 2 D) (1-D) (H_{AE}-H_{AB}),
\nonumber
\end{eqnarray}
where $H_{AE}$ is the entropy of $p \equiv P(A=E)$, ie.
\begin{equation}
H_{AE} = -p \mbox{ } log_2 p - (1-p) log_2 (1-p).
\end{equation}

\begin{figure}[h]
\begin{center}
\epsfxsize=9cm \epsfbox{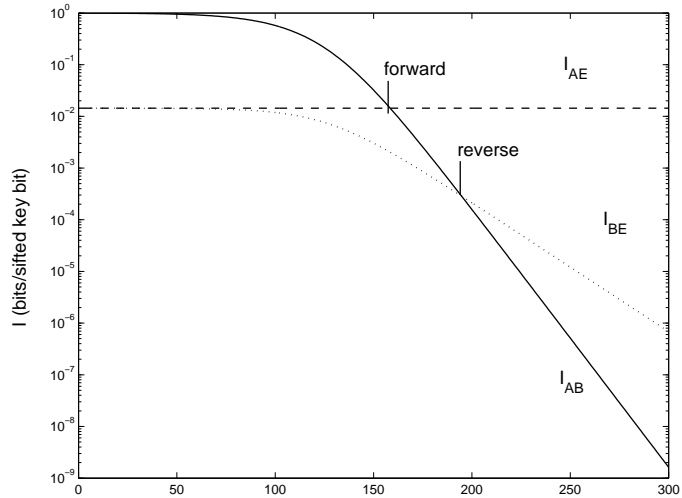} \caption{Mutual Information
with distance.  Forward and reverse mark the maximum secret key
distances using forward and reverse reconciliation (see text).}
\label{onelegI}
\end{center}
\end{figure}

Fig \ref{onelegI} shows the mutual information between Alice, Bob
and Eve, per bit of sifted key.  Eve's information about Alice's
sent bit stays constant.  Bob's only significantly changes in the
region where the signal to noise ratio becomes small, and quickly
becomes lower than Eve's. Roughly speaking, a key can be distilled
until the signal to noise ratio for Alice to Bob becomes small.
The main part of the noise is constant (dark counts), whilst the
main change of the signal is due to photon loss in the fibre.

An alternative method for distilling a key is called reverse
conciliation \cite{maurer, grangier}, and was previously used in
continuous variable quantum cryptography.  As we shall see, this
allows secure keys at longer distances than the conventional
forward reconciliation.  The idea is to do as before, only
centering the protocol on Bob rather than on Alice. This involves
one way communication from Bob to Alice, and gives a key rate of
\begin{equation}
Rate_{BA} = P(total) (I_{BA} - I_{BE}),
\end{equation}
where $I_{BA}=I_{AB}$.

The optimal eavesdropping attack for the reverse reconciliation is
the same as that for the forward reconciliation \cite{Yamamoto},
and for perfect sources and detectors gives Eve the same
information about Bob's bit as she learns about Alice's.  With
imperfect detectors, Eve will know less about Bob's bit than she
does about Alice, since the dark counts will introduce noise into
her knowledge.  Thus Alice and Bob can perform secure cryptography
at longer distances using such a protocol.

To calculate $I_{BE}$, first see that Eve's optimal strategy is
first to make a non-demolition measurement on the photon at the
end of the channel, to see whether or not it has passed the many
km of fibre (probability ($t$)), without destroying its quantum
state. If it has not ($1-t$), the only way Bob can see anything is
if he has a dark count, in which case Eve knows nothing about
Bob's bit, and $I_{BE}=0$.

If the photon has passed the channel, in order for Bob to have a
detection his detector must either detect the photon ($\eta$), or
have a dark count ($(1-\eta) 2 D$).  There must also be no dark
count in the other detector ($(1-D)$), and Alice and Bob must
measure in the same basis ($\frac{1}{2}$).  To get the fraction of
bits upon which Eve has some information, $P(photonpass)$, we
divide this by $P(total)$, giving
\begin{equation}
P(photonpass)  = \frac{t(\eta + (1-\eta) 2D)}{t \eta + (1-t\eta)
2D}.
\end{equation}

When the photon did indeed pass the channel, Eve's visibility is
given by using the imperfect $V_{opt}$ to eavesdrop, which in the
absence of dark counts would give her a visibility
$V_{BE}=\sqrt{1-V_{opt}^2}$.  With dark counts, Eve's visibility
is less, since she does not know if Bob detects a photon ($\eta$),
or a dark count ($(1-\eta)2D$).  The visibility becomes
\begin{equation}
V_{BE}(photonpass) = \frac{\eta \sqrt{1-V_{opt}^2}}{\eta+(1-\eta)2
D}.
\end{equation}
Following from this, the probability that Eve and Bob have the
same bit when the photon passed the channel
\begin{equation}
P(B=E|photonpass) = \frac{1}{2}(1+V_{BE}(photonpass)).
\end{equation}

Now we have everything necessary to write down $I_{BE}$.
\begin{equation}
\label{Ibe}
I_{BE} = P(photonpass) (1-H(P(B=E|photonpass))),
\end{equation}
and is plotted as the dotted line in fig \ref{onelegI}.

\begin{figure}[h]
\begin{center}
\epsfxsize=9cm \epsfbox{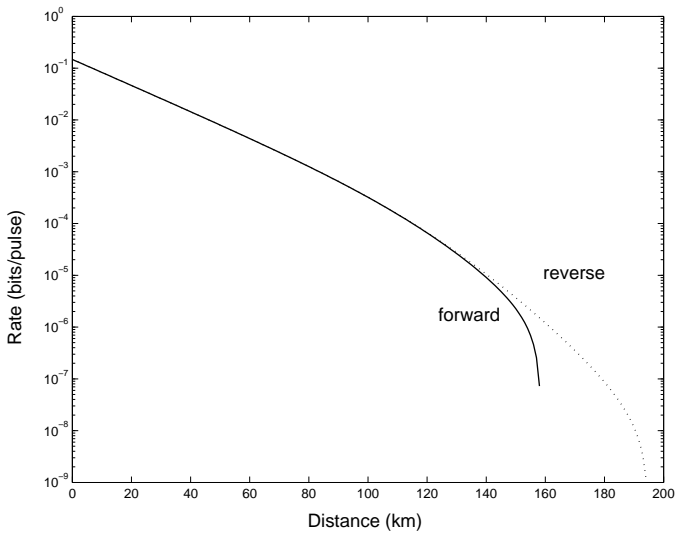} \caption{Secret key rate
with distance.  Solid line: forward reconciliation; dotted line:
reverse reconciliation.} \label{onelegrate}
\end{center}
\end{figure}

Fig \ref{onelegrate} shows the secret key rate which can be
distilled at various different distances.  For small distances, it
drops exponentially with increasing distance due to the decrease
in the sifted key length. For larger distances, it drops even more
quickly to zero at some finite distance. Reverse reconciliation
only gives a small advantage over forward reconciliation, and so
we shall ignore this difference in our relay calculations.

Another possibility to distill a secret key at longer distances
would be to use two way communication, ie. advantage distillation
\cite{advantage}. However the efficiency of all known protocols is
so low that we shall neglect them.

\section{The Quantum Relay}

With a single section of channel, secure key creation over
increasing distances becomes impossible essentially because the
signal at Bob's detector decreases, whereas the dark counts remain
constant.  The first step to combatting this is to split the
channel into two sections, and place a (perfect) singlet source
(making $\frac{1}{\sqrt{2}}(\ket{01}-\ket{10})$) in the middle.
One expects a key to be possible until the signal to noise ratio
in each detector is similar to what it was with just one section.
Since each photon only travels down half of the channel, this
means that the maximum total distance roughly doubles.

Since we are no longer sending a photon from Alice to Bob we need
a new protocol.  We use BBM92 \cite{BBM92} (a development of
Ekert's protocol \cite{Ekert}), which is conceptually similar to
BB84. Alice and Bob each measure in one of the two bases of the
BB84 protocol, and know that the results should be perfectly
anti-correlated if they measure in the same basis.  They then
perform sifting, error correction and privacy amplification as for
the BB84 protocol.

The next step is to split the channel into three sections, and use
a teleportation protocol.  This should give us a total distance
about three times that of the single section protocol.  After this
four sections is entanglement swapping, and we can keep adding
sections, hoping to get an n-fold increase in distance for
n-sections. However adding more sections also adds more sources of
noise, which will make adding more sections counter-productive
after some optimal number of sections. This limits the maximum
possible distance of this method.

To see these effects quantitatively, we shall begin by calculating
Bob's visibility of Alice's qubit.  We shall look at increasing
the distance and more sections.  Then we shall turn to
eavesdropping, and the mutual information between Alice, Bob and
Eve.  Next we shall look at the key rates, and give a simple
approximate formula for the maximum possible distance at which a
key can be distilled for different numbers of sections.  Finally
we shall examine the experimentally tunable tradeoff between
increased detector efficiency and more dark counts, and see how
this affects the key rates and maximum possible distances.

\section{Visibility of the Quantum Relay}

For the visibility between Alice and Bob, we begin with two
sections. To have a sifted key bit, both Alice and Bob must have
exactly one click.  Since the source is perfect, these
probabilities are independent.  If the source is in the middle of
the channel they are both equal to
\begin{equation}
p_2 = (t^{\frac{1}{2}} \eta + (1-t^{\frac{1}{2}} \eta) 2 D) (1-D),
\end{equation}
 where $t^{\frac{1}{2}}$ is the probability for a photon to
pass half of the channel.  Thus the probability of a sifted key
bit is
\begin{equation}
P(total) = \frac{1}{2} p_2^2.
\end{equation}
The signal, when Alice and Bob both receive something noiseless,
is
\begin{equation}
P(signal) = \frac{1}{2} V_{2opt} \left( t^{\frac{1}{2}} \eta (1-D)
\right)^2,
\end{equation}
where $V_{2opt}$ is the optical visibility across the whole length
of the channel.

For three sections, we use teleportation, with a linear optics
Bell measurement.  It is not possible to distinguish the four Bell
states using linear optics \cite{Calsamiglia}: the best one can do
is to have three outcomes: the first Bell state, the second Bell
state, and "not the first or second Bell states".  For example, we
may encode our qubits as single photon time bins, arriving {\it
now} encoding a $\ket{0}$, and arriving {\it later} encoding a
$\ket{1}$.  If our two qubits are in different fibres, we may
combine them on a beam-splitter and then use detectors which
measure the time of arrival of the photon in order to perform the
Bell measurement. An input state
$\frac{1}{\sqrt{2}}(\ket{01}-\ket{10})$ gives two photons on
different sides and at different times after the beam-splitter,
whilst $\frac{1}{\sqrt{2}}(\ket{01}+\ket{10})$ gives two photons
at different times on the same side.  The other two states,
$\frac{1}{\sqrt{2}}(\ket{00} \pm \ket{11})$, give two photons on
the same side at the same time, and thus we only see one photon.

Since the Bell measurement only works half the time for
uncorrelated inputs (such as we have here), the noiseless signal
is now
\begin{equation}
P(signal) = \frac{1}{2} V_{3opt} \left( t^{\frac{1}{3}} \eta (1-D)
\right)^3 \frac{1}{2}.
\end{equation}
The total signal is more complicated than before.  The chance of
Bob getting one count at his detector is similar to before,
\begin{equation}
p_3=(t^{\frac{1}{3}} \eta + (1-t^{\frac{1}{3}} \eta) 2 D) (1-D).
\end{equation}
However the chance of accepting the clicks at the Bell measurement
is slightly reduced from $p_3^2$, since half the time when the two
photons both trigger a detector, they are both triggering the same
detector at the same time, and we only see one detection.  So we
subtract this probability: ($-\frac{1}{2}(t^{\frac{1}{3}} \eta)^2
(1-D)^2$). If this happens (ie. both photons are "detected", but
we only see one detection event), we can still see a coincidence
if there is a dark count in another detector. Since there are two
other detectors where the dark count would give us the same result
as a Bell state, this has probability ($+2 D
\frac{1}{2}(t^{\frac{1}{3}} \eta)^2 (1-D)^2$). This gives a total
probability
\begin{equation}
P(total) = \frac{1}{2} p_3 \left( p_3^2 - (1-2D) \frac{1}{2} (\eta
t^{\frac{1}{3}})^2 (1-D)^2 \right).
\end{equation}

We now have all the elements to generalise this to n sections.
\begin{equation}
P(signal) = \left( \frac{1}{2} \right)^{\lfloor \frac{n+1}{2}
\rfloor} V_{nopt} \left( t^{\frac{1}{n}} \eta (1-D) \right)^n,
\end{equation}
\begin{equation}
P(total) = \frac{1}{2} p_n^{n-2\lfloor \frac{n-1}{2} \rfloor}
\left( p_n^2 - (1-2D) \frac{1}{2} (\eta t^{\frac{1}{n}})^2 (1-D)^2
\right)^{\lfloor \frac{n-1}{2} \rfloor} ,
\end{equation}
\begin{equation}
p_n=(t^{\frac{1}{n}} \eta + (1-t^{\frac{1}{n}} \eta) 2 D) (1-D),
\end{equation}
where $\lfloor x \rfloor$ is $x$ rounded down to the nearest
integer.

\begin{figure}[h]
\begin{center}
\epsfxsize=9cm \epsfbox{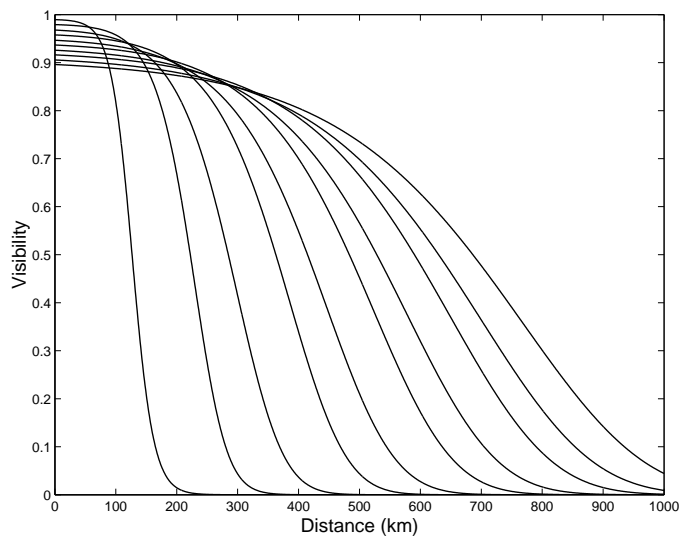} \caption{Bob's visibility of
Alice's qubit for a channel split into 1..10 sections.}
\label{nlegV}
\end{center}
\end{figure}

In Fig \ref{nlegV} we show the ten visibility curves for splitting
a channel of increasing distance into 1..10 sections.  The
leftmost curve is for 1 section, then moving to the right we pass
through 2, 3, and so on up to 10.  We take $V_{nopt}=V_{opt}^n$,
which is why the visibility at $0$ distance decreases with
increasing sections.

Using more sections allows us to increase the visibility at long
distances, and for low visibilities we have the approximately
linear increase in distance with the number of sections that we
mentioned before.  There is a limit to this: however many relays
we use we shall never get a better visibility than $0.9$ at more
than $300km$. This is because we are introducing more detectors
and hence more dark counts with more sections. Moving from an even
number of sections to an odd is not as useful as the other way
around since we have to add a Bell measurement, which is noisy
since it only works for half of the Bell states.

\section{The Quantum Relay For Cryptography}

To see how the quantum repeater helps cryptography, we have to
calculate the various mutual informations.  Between Alice and Bob,
this is simple: for any number of sections
\begin{equation}
I_{AB} = 1 - H \left( \frac{1}{2} + \frac{1}{2}
\frac{P(signal)}{P(total)} \right).
\end{equation}

To calculate Eve's information, we begin with two sections.  For
her optimal strategy, we view the protocol as if it is BB84, and
assume the same attack as before. This was shown to be the optimal
individual attack in \cite{Yamamoto}.  The exact equation for
Eve's knowledge of Alice's bit is more complicated than before,
because of the dark counts in Alice's detectors the value of which
she knows nothing about.  In this respect it is similar to Eve's
knowledge of Bob in the BB84 case (equation (\ref{Ibe})).
\begin{equation}
I_{AE} = P_2(photonpass) \left( 1 - H \left( \frac{1}{2} (1 +
V_{AE,2}(photonpass)) \right) \right),
\end{equation}
where
\begin{equation}
P_2(photonpass) = \frac{t^{\frac{1}{2}}(\eta + (1-\eta)
2D)}{t^{\frac{1}{2}} \eta + (1-t^{\frac{1}{2}}\eta) 2D},
\end{equation}
\begin{equation}
V_{AE,2} = \frac{\eta \sqrt{1-V_{2opt}^2}}{\eta+(1-\eta)2D}.
\end{equation}

By symmetry, $I_{BE} = I_{AE}$.

For three sections, Eve can use the errors due to the dark counts
at the Bell measurements as well as the optical visibility to
eavesdrop. The visibility between Alice and Bob of these errors is
given by
\begin{equation}
V_{AB}^E =  V_{3opt} \frac{\frac{1}{2} (t^{\frac{1}{3}}
\eta)^2}{p_3^2 - (1-2D) \frac{1}{2} (\eta t^{\frac{1}{3}})^2 } .
\end{equation}
Eve's visibility of Alice's bit is given by $\sqrt{1- \left(
V_{AB}^E \right)^2 }$, and so
\begin{equation}
I_{AE} = 1-H \left( \frac{1}{2} + \frac{1}{2} \sqrt{1- \left(
V_{AB}^E \right)^2 } \right).
\end{equation}
Eve's visibility of Bob's bit is
\begin{equation} I_{BE} = P_3(photonpass) \left( 1 -
H \left( \frac{1}{2} (1+ V_{AE,n} \right) \right),
\end{equation}
where
\begin{equation}
P_n(photonpass) = \frac{t^{\frac{1}{n}}(\eta + (1-\eta)
2D)}{t^{\frac{1}{n}} \eta + (1-t^{\frac{1}{n}}\eta) 2D},
\end{equation}
\begin{equation}
V_{AE,n} = \frac{\eta \sqrt{1- \left( V_{AB}^E \right)^2
}}{\eta+(1-\eta)2D}.
\end{equation}

We can now write down the equations for dividing the channel into
n equal sections.

\begin{equation}
V_{AB}^E = V_{nopt} \left( \frac{ \frac{1}{2}(t^{\frac{1}{n}}
\eta)^{2 } }{ p_n^2 - (1-2D) \frac{1}{2} (\eta t^{\frac{1}{n}})^2
} \right)^{\lfloor \frac{n-1}{2} \rfloor}  .
\end{equation}

\begin{equation}
I_{AE} = \left\{ \begin{array}{ll} 1-H \left( \frac{1}{2} +
\frac{1}{2} \sqrt{1- \left( V_{AB}^E \right)^2 } \right) &
\mbox{for
 odd n} \\ I_{BE} & \mbox{for even n}
\end{array} \right.
\end{equation}
\begin{equation} I_{BE} = P_n(photonpass) \left( 1 -
H \left( \frac{1}{2} (1+ V_{AE,n}) \right) \right).
\end{equation}

\begin{figure}[h]
\begin{center}
\epsfxsize=9cm \epsfbox{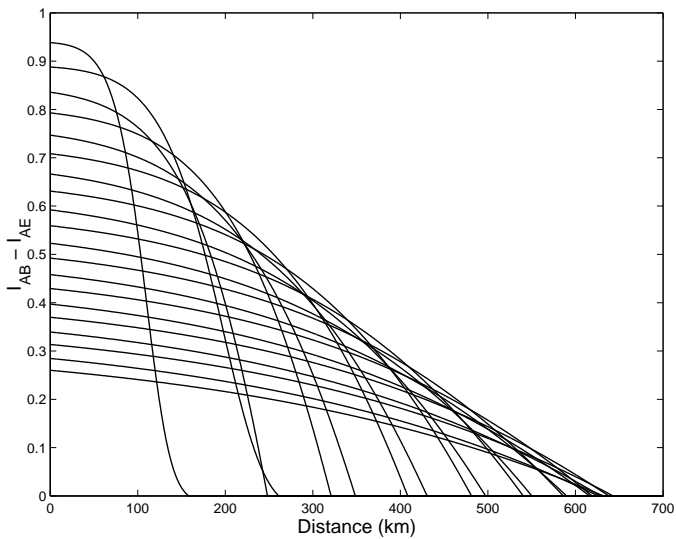} \caption{$I_{AB}-I_{AE}$ for
the quantum relay with 1..20 sections.} \label{nlegI}
\end{center}
\end{figure}

As before, the secret key rate is $P(total) (I_{AB} - I_{AE})$.
Fig \ref{nlegI} shows the difference in the mutual informations
for various distances when splitting the channel into 1..20
sections. In order to simplify the calculations, we have used the
bound
\begin{equation}
P(A=E) \le \frac{1}{2} + \frac{1}{2} \sqrt{1- \left( V_{AB}^E
\right)^2}.
\end{equation}
This bound comes from assuming (for Eve's purposes) that there are
no dark counts, giving her more information than she really has.

More sections give us a significant improvement over a single
section.  There is a limit to the distance we can achieve with the
relay as the benefits of adding more sections are negated by the
extra dark counts and losses introduced. Moving from an even
number of sections to an odd number does not gain very much
because we introduce an extra Bell measurement, whose dark count
noise can be used by Eve for eavesdropping.

\begin{figure}[h]
\begin{center}
\epsfxsize=9cm \epsfbox{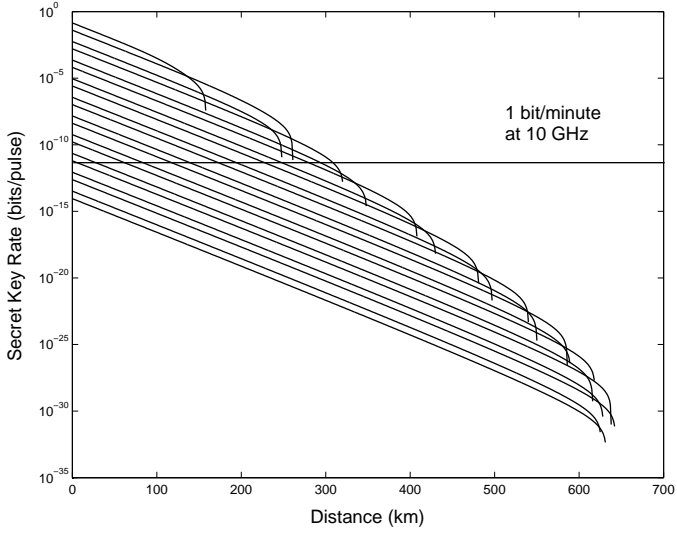} \caption{Distilled key rete
for the quantum relay with 1..20 sections.} \label{nlegrate}
\end{center}
\end{figure}

Fig \ref{nlegrate} shows the secret key rate for several sections.
Adding more sections does not increase the rate at short
distances, but rather increases the maximum distance at which a
secret key can be distilled.  There are two limits to the maximum
number of useful sections.  The more fundamental one is that after
around 18 sections we are introducing too much noise and so no
longer gaining anything.  The more practical one is that even if
we pulse the system at 10 GHz, the key rate with more than four
sections will be too low to be of any use.  We have drawn a line
where the secret key will be generated at 1 bit/minute if pulsed
at 10 GHz.  This problem will be of even more consequence in the
laboratory, where the sources are not perfect and the rate will be
lower still.  Although there is a fundamental limit of around
$650$ km using our quantum relay assumptions, the more important
limit is when the losses make the key rate too low to be of use.

\section{The maximum achievable distance}

\begin{figure}[h]
\begin{center}
\epsfxsize=9cm \epsfbox{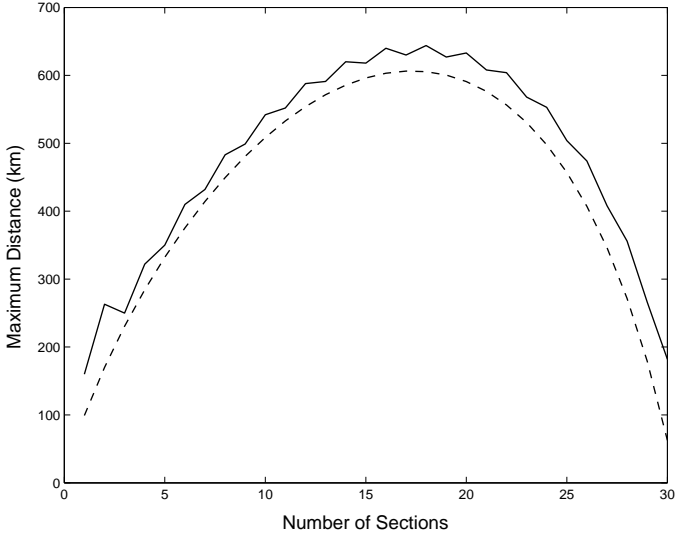} \caption{The maximum distance
at which a secret key can be generated for 1..30 sections. Solid
line: calculated maximum distance; dashed line: estimate (see
below).} \label{maxdist}
\end{center}
\end{figure}

Fig \ref{maxdist} shows the maximum distance at which we can
extract a secret key when we split the channel into increasing
numbers of sections.  For the parameters used here, the maximum is
for 18 sections and allows a secret key at around 650 km.

This behaviour can be understood by a much simpler formula. First
assume that all the noise in the system, including the dark counts
in Alice and Bob's laboratories, is available for eavesdropping.
This will be a good approximation for large numbers of relays,
since most of the detectors will be Eve's possession. Then Eve's
visibility of Alice's bit is $\sqrt{1-V_{AB}^2}$. In order to
distill a secret key, Bob must have a better visibility of Alice's
bit than Eve, ie.
\begin{equation}
V_{AB} > \sqrt{1-V_{AB}^2}.
\end{equation}
Solving this gives
\begin{equation}
\label{maxV}
V_{AB} > \frac{1}{\sqrt{2}}.
\end{equation}
To simplify the calculation of the visibility, we shall assume
that $(1-D) \simeq 1$, and that for long distances $(1 -
t^{\frac{1}{n}} \eta) \simeq 1$.  If we imagine for a minute that
the Bell measurement is perfect then
\begin{eqnarray}
V_{AB} & \simeq & \frac{\frac{1}{2} V_{nopt} t \eta^n}{\frac{1}{2}
(t^{\frac{1}{n}} \eta +
 2 D)^n} \\
& \simeq & \left( \frac{V_{opt}}{1 + \frac{2 D}{t^{\frac{1}{2}}
\eta }} \right)^n.
\end{eqnarray}
Together with equation (\ref{maxV}) this gives
\begin{equation}
d_{max}(\mbox{perfect bell measurement}) \simeq \frac{10
n}{\alpha} log_{10} \left( \frac{\eta}{2 D} ( 2^{\frac{1}{2n}}
V_{opt}-1) \right).
\end{equation}
We just need to modify this for the effect of imperfect Bell
measurements.  These replace the term $(t^{\frac{1}{n}} \eta +
 2 D)^2$ by $(t^{\frac{1}{n}} \eta +
 2 D)^2 -  \frac{1}{2}(1-2D)(t^{\frac{1}{n}} \eta)^2$.  Assuming
 that $D \ll 1$, and that $t^{\frac{1}{n}} \eta \gg
 2 D$, we have
\begin{equation}
 (t^{\frac{1}{n}} \eta +
 2 D)^2 -  \frac{1}{2}(t^{\frac{1}{n}} \eta)^2 \simeq \frac{1}{2}
 (t^{\frac{1}{n}} \eta + 4 D)^2.
\end{equation}
We see that the effect of the imperfect measurements is roughly to
double the effective dark counts.  Thus,
\begin{equation}
\label{dmax}
 d_{max} \simeq \frac{10 n}{\alpha} log_{10} \left(
\frac{\eta}{4 D} ( 2^{\frac{1}{2n}} V_{opt}-1) \right).
\end{equation}
The dashed line in Fig \ref{maxdist} plots this function, which is
similar to the results obtained using the more precise equations.

\section{Detector Efficiency vs. Dark Counts}
\label{darkvefficiency}

\begin{figure}[h]
\begin{center}
\epsfxsize=9cm \epsfbox{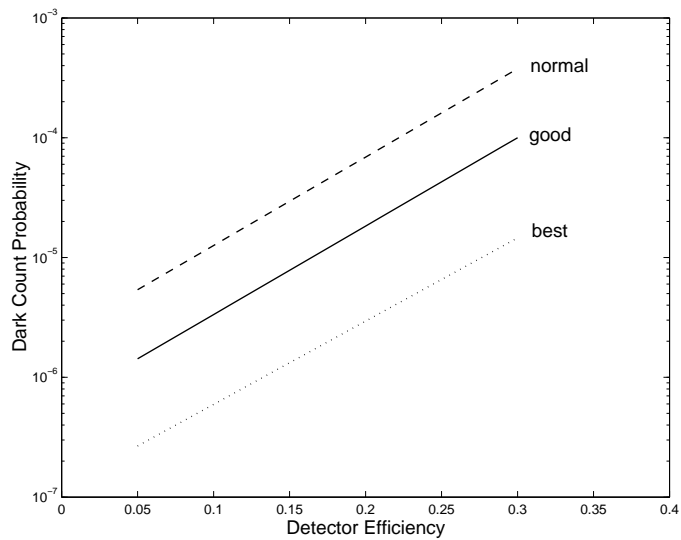} \caption{The tradeoff
between efficiency and dark counts for high quality single photon
detectors at optical wavelength.} \label{detector}
\end{center}
\end{figure}

A key part in any cryptosystem is the quality of the detectors. We
would like the detector efficiency to be as high as possible, so
that the efficiency of the relay is not too low.  We would also
like the dark count probability to be as low as possible, to
reduce the noise and allow us to make a secret key at long
distances.  In general these two aims pull in different
directions: increasing the efficiency also increases the dark
counts.  We have to compromise to find the best detector for our
setup.  Fig \ref{detector} shows the efficiency/dark count
tradeoff for high quality InGaAs APD detectors operating at
$-50^{\circ} C$ at telecom wavelength, 1550 nm \cite{Gregoire}. We
shall

We shall concentrate on the "good" detectors, with line
$d=Ae^{B\eta}$, with $A=6.1 \times 10^{-7}$ and $B=17$. This is
for detectors with a quality roughly half way between the normal
ones ($A=2.3 \times 10^{-6}$, $B=17$), and the best ones found to
date ($A=1.2 \times 10^{-7}$, $B=16$), which are quite rare. The
parameters $d=10^{-4}$, $\eta=0.3$ which we have used in this
paper lie on the "good" line.

In order to get the maximal possible distance at which a key can
be distilled, we see from equation (\ref{dmax}) that we need to
maximize $\frac{\eta}{D}$.  The best detector is the one with the
smallest possible dark counts, ie. $\eta=0.05$, $d\simeq 10^{-6}$.

\begin{figure}[h]
\begin{center}
\epsfxsize=9cm \epsfbox{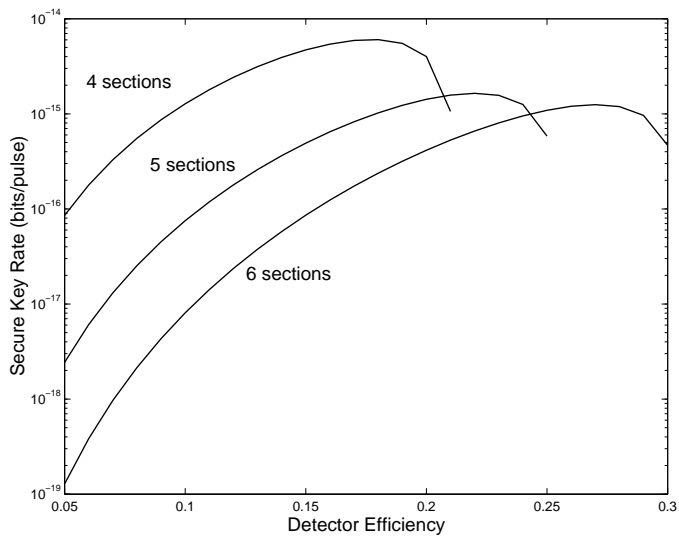} \caption{Secret key rates at
400 km for 4, 5 and 6 sections using the "good" detectors in fig
\ref{detector}.} \label{dvseta}
\end{center}
\end{figure}

The problem with such detectors is that the efficiency is so low
that the rate of secret key generation becomes very low.  For a
fixed distance, it is usually better to take detectors with better
efficiency.  In fig \ref{dvseta} we plot the secret key rates at
400 km of relays with 4..6 sections with the good detectors of Fig
\ref{detector}.  The rates for 1..3 sections are $0$.  The rates
are all low since 400 km of fibre has a transmission probability
of $10^{-10}$, and we also need a coincidence between 4, 5 or 6
photons.

We see that the best rates are attained when using the fewest
sections of relay possible, in this case four.  For a fixed number
of sections, the rates generally increase with increasing
efficiency (and hence increasing dark counts), until just before
the dark counts become large and forbid us from making a secret
key.  For this distance we should take an efficiency around
$0.18$.

The main lesson here is that whilst we need detectors with low
dark counts to allow us a non-zero key rate at longer distances,
it is equally important for the quantum relay to find detectors
with a high efficiency.  Such detectors would allow a higher key
rate at long distances, extending the range of practical quantum
cryptography.

\section{Conclusion}

The quantum relay significantly improves the distance at which
quantum cryptography is possible.  With the physical assumptions
made here, one can go about four times as far with a multi-section
relay as one can with a single section protocol.  The distance of
the relay is not unlimited because the noise added by additional
sections counteracts the potential gain. We hope that even with
present technology one may be able to go to two or three times the
single-section distance by using an arrangement with only three
\cite{222} or four sections.

We have also seen that for the quantum relay to be useful, we need
detectors with a high efficiency.  This is because at long
distances we already have lots of loss in the fibre.  If we
introduce too many more losses by adding detectors, we are left
with a key rate so small as to be effectively $0$.

One configuration which is better than those we have considered
here uses high efficiency (and hence high dark count probability)
detectors for Alice and Bob, and detectors with much lower dark
counts (and so lower efficiency) in the middle of the relay. This
is useful since the dark counts in Alice and Bob's labs cannot be
used by Eve, and so we boost the count rate whilst keeping good
security.

This is a small step towards quantum cryptography on a national
scale.  Improved sources and detectors will increase the distance,
but either coherent manipulation of large numbers of qubits or
satellites will be required to bring quantum cryptography to the
world scale. We hope that one day, using small sources and
detectors, a quantum cryptographic modem will form a part of
everyone's personal computer.

 \vline

{\bf Acknowledgements} We thank A. Acin, G. Ribordy, V. Scarani,
W. Tittel and H. Zbinden for helpful discussions.

We acknowledge funding by the Swiss NCCR, "Quantum Photonics" and
the European IST project RESQ.

\end{document}